# Exchange stiffness proportional to power of magnetization in permalloy co-doped with Mo and Cu

Shiho Nakamura, Nobuyuki Umetsu, Michael Quinsat, and Masaki Kado

Institute of Memory Technology Research and Development, Kioxia Corporation

3-13-1, Moriya-cho, Kanagawa-ku, Yokohama, 221-0022, Japan

## Abstract

The exchange stiffness of magnetic materials is one of the essential parameters governing magnetic texture and its dynamics in magnetic devices. The effect of single-element doping on exchange stiffness has been investigated for several doping elements for permalloy (NiFe alloy), a soft magnetic material whose soft magnetic properties can be controlled by doping. However, the impact of more practical multi-element doping on the exchange stiffness of permalloy is unknown. This study investigates the typical magnetic properties, including exchange stiffness, of permalloy systematically co-doped with Mo and Cu using broadband ferromagnetic resonance spectroscopy. We find that the exchange stiffness, which decreases with increasing doping levels, is proportional to a power of magnetization, which also decreases with increasing doping levels. The magnetization, $M_s$, dependence of the exchange stiffness constant, $A$, of all the investigated samples, irrespective of the doping levels of each element, lies on a single curve expressed as $A \propto M_s^n$ with exponent $n$ close to 2. This empirical power-law relationship provides a guideline for predicting unknown exchange stiffness in non-magnetic element-doped permalloy systems.

## 1. Introduction

Exchange stiffness represents the amount of energy difference associated with a spatial change in spins (magnetic moments) in a non-uniformly magnetized ferromagnet, reflecting the strength of the interaction between spins in the material [1-3]. It determines magnetic textures, including magnetic domain [1], domain wall width [3-7], magnetic skyrmions [8,9], and vortexes [10,11], as well as their dynamics [12-16], together with magnetization and magnetic anisotropy. Therefore, it is one of the essential parameters to simulate and then predict the performance of various magnetic devices such as magnetic memories and logics. Despite its indispensability, there is far less information compared to the other basic parameters of magnetic materials, magnetization and magnetic anisotropy. One of the reasons for this is the lack of simple methods to measure exchange stiffness constant $A$. An efficient way to measure $A$ is to excite spin waves and measure their spectra. Excitation methods include neutron scattering [17,18], Brillouin light scattering (BLS) using photons in the visible range [19-22], and ferromagnetic resonance (FMR) using microwaves of fixed [23,24] or broadband frequencies [22, 25-27]. In particular, broadband FMR with microstrip lines or coplanar waveguides, which have become readily available due to recent advances in high-frequency technology, is a relatively accessible and sensitive method that can also measure other parameters including magnetic damping constant.

Permalloy, based on NiFe alloys, is a well-known typical soft magnetic material [28-31]. In particular, NiFe alloys in combination with several doping elements, such as Mo and Cu, are widely used in practical applications due to their ultra-high magnetic permeability [29-31]. Changes in fundamental physical properties of permalloy, including magnetization and exchange stiffness, with doping, have been investigated for single-element doping such as Pt, Cu, V, and Ta by means of spin-wave measurements using FMR and BLS methods, and *ab initio* calculations [22,25,27,32-33]. However, to the best of our knowledge, there are no reports on the exchange stiffness of multi-element doped systems, which are closer to practical materials.

In this study, we report on the effect of Mo and Cu co-doping on the exchange stiffness of permalloy along with the effect on other typical magnetic properties including magnetization, *g*-factor, and Gilbert damping constant. The composition ratio of NiFe (Py) is 80:20 at. % and the doping level of Mo ranges from 0 to 9.4 at. % in Mo/Py, and that of Cu from 0 to 11.7 at. % in Cu/Py. Broadband FMR method is used to evaluate the parameters. We find the dependence of exchange stiffness on saturation magnetization, $M_s$, exhibits a relationship, $A \propto M_s^n$, with the exponent $n$ is 2.24 ± 0.04, for all combinations of Mo and Cu doping levels investigated. We also investigate the magnetization dependence of exchange stiffness in single-element doped systems from literatures, and find that the power law relationship seems to be a general trend for non-magnetic element doped permalloy systems.

## 2. Experimental

The investigated Py-Mo-Cu samples were formed by alternating deposition of sub-atomic layers of the constituent elements in order to systematically control the composition. A magnetron sputter deposition system with base pressure less than $3\times10^{-6}$ Pa was used for deposition, and Si wafers with a thermal oxidation layer were used as substrates. The film stacking procedures were [Fe (0.11 nm) / Ni (0.35 nm) / Cu ($n \times$ 0.02 nm, $n$ = 0~3) / Mo ($m \times$ 0.01 nm, $m$ = 0~4)] × 110 cycles and [Fe (0.11 nm) / Ni (0.35 nm) / Cu (4 × 0.02 nm)] × 110 cycles. All the films were capped with Ta (3 nm) and $SiO_2$ (10 nm) to protect surface oxidation. The thicknesses of the obtained Py-Mo-Cu film ranged from 49.7 nm to 62.5 nm, depending on the doping level. The composition of the films was determined by X-ray florescence (XRF) measurements with fundamental parameter analysis [34]. A typical XRF spectrum of the film is shown in Fig. 1. The average Py concentration in the 21 samples was 80.2 at. % for Ni and 19.8 at. % for Fe, with a standard deviation of 0.7 at. % for both Ni and Fe in the 21 samples. An excellent linear relationship between the measured doping concentration and the number of sub-atomic layers of dopant was confirmed. Table 1 summarizes the film composition.

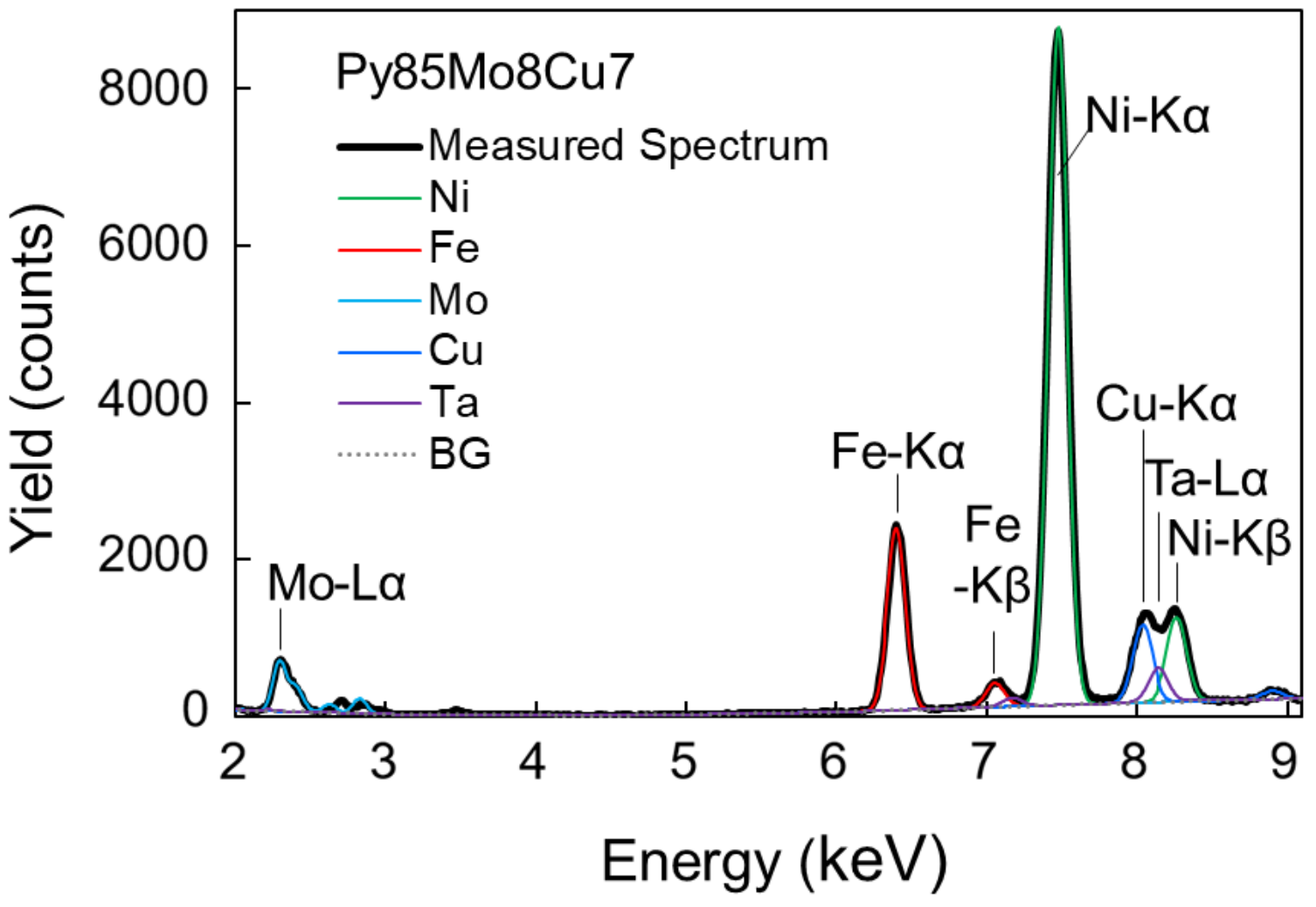

Figure 1. X-ray florescence spectra of Py co-doped with 9.4 at. % Mo/Py and 9.4 at. % Cu/Py (Py85Mo8Cu7). The thick black curves represent the measurement results and the thin colored curves represent each component analyzed. BG indicates the background signal.

Here, linear fit values were used as nominal values for Cu and Mo concentrations in Cu/Py and Mo/Py, respectively. The difference between these nominal values and measured values is negligible, about ±0.1 at. % in average and a maximum of 0.4 at. % for highest concentration of Cu. Therefore, we used the nominal values in all the manuscript. The X-ray diffraction patterns for each film showed a single diffraction peak corresponding to (111) reflections of face-centered cubic NiFe alloy, suggesting that the elements are well alloyed and the films are polycrystalline with <111> preferred crystal orientation. The lattice constant of undoped Py was 0.3546 nm, which is in good agreement with that of the bulk value of Py [17]. The presence of Cu dopants hardly changed the lattice constant, as reported for a single Cu doping [22]. Whereas the addition of 9.4 % Mo increased it to 0.3581 nm, but the increase rate is as small as 1 %.

Typical magnetization curves measured using a vibrating sample magnetometer (VSM) are shown in Fig. 2. The coercive force for the magnetic field applied along easy-axis was less than 1 mT for all the compositions. Whereas for the hard-axis, the anisotropy field $H_k$ increased from 1.7 mT to 11.7 mT, depending on the doping levels, particularly affected by Mo doping as shown in Table 1. This uniaxial anisotropy must be caused by the induced magnetic anisotropy during deposition and by the non-zero-magnetostriction due to doping [30,35].

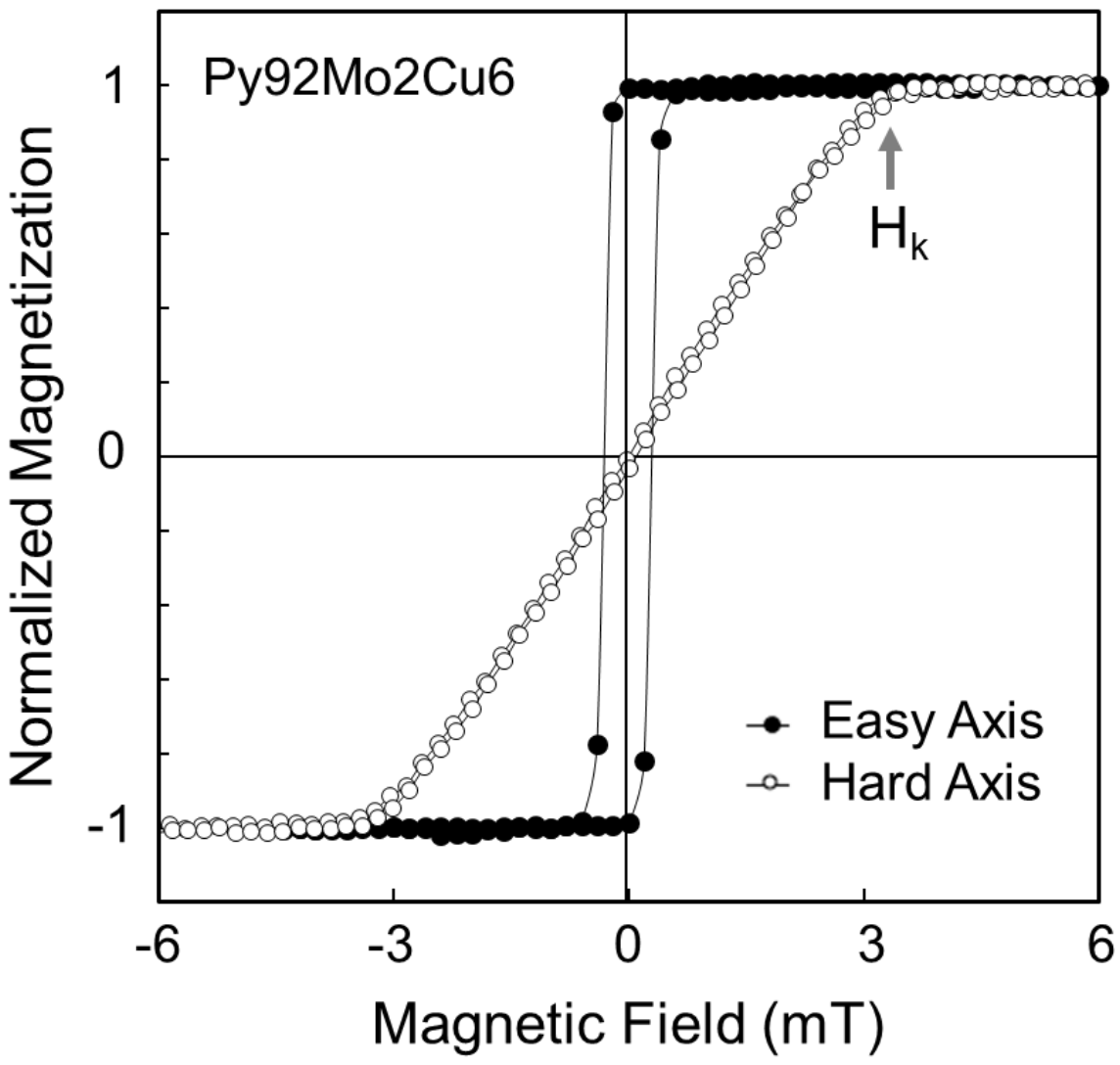

Figure 2. Magnetization curves of Py co-doped with 2.3 at. % Mo/Py and 5.9 at. % Cu/Py (Py92Mo2Cu6) for easy- and hard-axis.

FMR and perpendicularly standing spin wave (PSSW) were measured using the Broadband FMR method using a coplanar waveguide with microwave frequency range up to 40 GHz. The external field of up to 1 T was applied in the plane of the film perpendicular to the microwave field. Typical measurements of both FMR and PSSW modes are shown in the insets in Fig. 3 with the PSSW mode appearing at lower field. The spectra were fit with a sum of symmetrical and antisymmetrical Lorentzian derivatives in Equation (1) in Ref. 25. The field dependence of the extracted FMR and PSSW resonance frequencies are shown in Fig. 3, together with fits to the equation for films with in-plane uniaxial anisotropy,

$$f = \frac{\gamma\mu_0}{2\pi}\left[\left(H_{\mathrm{res}} + H^{a}_{\mathrm{k}(\varphi)} + \frac{2A}{\mu_0 M_{\mathrm{s}}}\left(\frac{p\pi}{d}\right)^2\right) \times \left(H_{\mathrm{res}} + H^{b}_{\mathrm{k}(\varphi)} + M_{\mathrm{s}} + \frac{2A}{\mu_0 M_{\mathrm{s}}}\left(\frac{p\pi}{d}\right)^2\right)\right]^{\frac{1}{2}}, \qquad (1)$$

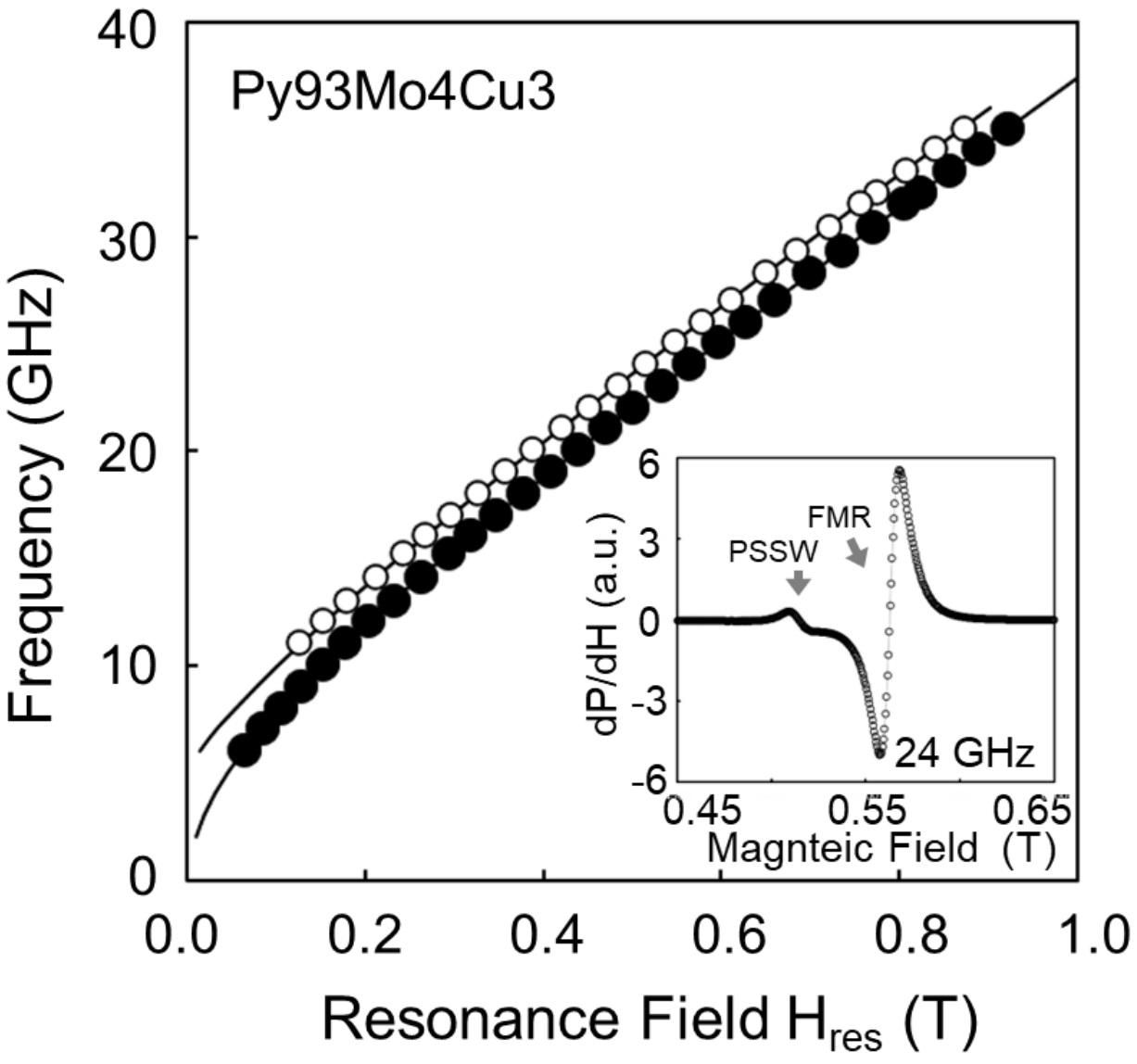

Figure 3. Frequency versus resonance magnetic field of the FMR mode (closed circles) and PSSW mode (open circles) for Py co-doped with 4.7 at. % Mo/Py and 2.9 at. % Cu/Py (Py93Mo4Cu3). The solid lines are fits to the Kittel equation. Inset shows the resonance curve of the same sample for frequency of 24 GHz.

where *f* is the frequency, $H_{\mathrm{res}}$ is the magnetic resonance field of either the FMR ($p$ = 0) or the PSSW mode ($p$ = 1), $\gamma$ is the gyromagnetic ratio, $\mu_0$ is the permeability of free space, $H^a_{\mathrm{k}(\varphi)}$ and $H^b_{\mathrm{k}(\varphi)}$ are the in-plane anisotropy fields expressed as $H^a_{\mathrm{k}(\varphi)} = -H_{\mathrm{k}}\cos(\pi - 2\varphi)$ and $H^b_{\mathrm{k}(\varphi)} = H_{\mathrm{k}}\cos^2\varphi$, $\varphi$ is the azimuthal angle of the applied field from the easy axis, and *d* is the thickness of the film [25,36-38]. When $\varphi$ is zero, $H^a_{\mathrm{k}(\varphi)}$ = $H^b_{\mathrm{k}(\varphi)}$ = $H_{\mathrm{k}}$. For $H_{\mathrm{k}}$, we used the values obtained from VSM measurements. From such fits, we extracted $M_{\mathrm{s}}$ and $\gamma$ from the FMR mode and then, extracted *A* from the PSSW mode. $\gamma$ gives the *g*-factor *g* from the relation $\gamma = 2\pi g\mu_{\mathrm{B}}/h$. The determined values of un-doped Py were $\mu_0 M_{\mathrm{s}}$ = 0.995 ± 0.028 T, *A* =11.7 ± 1.2 pJ/m, and *g* = 2.09 ± 0.01, which are consistent with the literatures [25,28,37,39]. Further, in order to derive Gilbert damping constant $\alpha$, the frequency-dependent FMR linewidth $\Delta H_{\mathrm{F}}$ of the films were fit using

$$\Delta H_{\mathrm{F}} = \Delta H_0 + \frac{4\pi\alpha}{\gamma} f, \qquad (2)$$

where $\Delta H_0$ is the inhomogeneous broadening [40,41]. The parameters extracted from the fits are listed in Table 1.

Table 1. Extracted values for $\mu_0 M_s$, $g$-factor, $A$, and $\alpha$ of Py-Mo-Cu alloys from FMR measurements.

| Cu/Py (at. %) | Mo/Py (at. %) | Composition | $H_k$ ($10^{-4}$ T) | $\mu_0 M_s$ (T) | $g$-factor | $A$ (pJ/m) | $\alpha$ ($10^{-3}$) |
|---|---|---|---|---|---|---|---|
| 0 | 0 | Py100 | 17 ± 2 | 0.995 ± 0.018 | 2.090 ± 0.012 | 11.72 ± 1.03 | 7.1 ± 0.2 |
| 0 | 2.3 | Py98Mo2 | 20 ± 2 | 0.833 ± 0.015 | 2.091 ± 0.011 | 7.91 ± 0.88 | 8.3 ± 0.2 |
| 0 | 4.7 | Py96Mo4 | 26 ± 2 | 0.680 ± 0.012 | 2.091 ± 0.010 | 4.82 ± 0.65 | 9.9 ± 0.2 |
| 0 | 7.0 | Py93Mo7 | 39 ± 2 | 0.533 ± 0.010 | 2.095 ± 0.009 | 2.70 ± 0.45 | 13.2 ± 0.2 |
| 0 | 9.4 | Py91Mo9 | 62 ± 2 | 0.393 ± 0.008 | 2.100 ± 0.009 | 1.63 ± 0.38 | 18.0 ± 0.3 |
| 2.9 | 0 | Py97Cu3 | 23 ± 2 | 0.943 ± 0.018 | 2.094 ± 0.012 | 10.74 ± 1.06 | 7.7 ± 0.3 |
| 2.9 | 2.3 | Py95Mo2Cu3 | 27 ± 2 | 0.785 ± 0.015 | 2.094 ± 0.011 | 6.75 ± 0.78 | 9.0 ± 0.2 |
| 2.9 | 4.7 | Py93Mo4Cu3 | 33 ± 2 | 0.632 ± 0.012 | 2.095 ± 0.010 | 4.11 ± 0.57 | 10.8 ± 0.2 |
| 2.9 | 7.0 | Py91Mo6Cu3 | 50 ± 2 | 0.490 ± 0.010 | 2.097 ± 0.009 | 2.47 ± 0.43 | 15.1 ± 0.4 |
| 2.9 | 9.4 | Py89Mo8Cu3 | 85 ± 2 | 0.352 ± 0.007 | 2.103 ± 0.008 | 1.44 ± 0.30 | 18.4 ± 0.5 |
| 5.9 | 0 | Py94Cu6 | 25 ± 2 | 0.894 ± 0.017 | 2.095 ± 0.011 | 9.36 ± 0.98 | 7.8 ± 0.2 |
| 5.9 | 2.3 | Py92Mo2Cu6 | 31 ± 2 | 0.737 ± 0.014 | 2.096 ± 0.011 | 5.92 ± 0.74 | 9.1 ± 0.2 |
| 5.9 | 4.7 | Py91Mo4Cu5 | 43 ± 2 | 0.593 ± 0.011 | 2.094 ± 0.010 | 3.63 ± 0.59 | 12.0 ± 0.3 |
| 5.9 | 7.0 | Py89Mo6Cu5 | 57 ± 2 | 0.446 ± 0.009 | 2.100 ± 0.009 | 2.04 ± 0.38 | 15.7 ± 0.4 |
| 5.9 | 9.4 | Py87Mo8Cu5 | 95 ± 2 | 0.309 ± 0.007 | 2.105 ± 0.009 | 1.20 ± 0.29 | 19.6 ± 0.6 |
| 8.8 | 0 | Py92Cu8 | 31 ± 2 | 0.852 ± 0.016 | 2.094 ± 0.012 | 8.31 ± 0.96 | 7.7 ± 0.2 |
| 8.8 | 2.3 | Py90Mo2Cu8 | 38 ± 2 | 0.693 ± 0.013 | 2.096 ± 0.010 | 5.15 ± 0.77 | 10.2 ± 0.2 |
| 8.8 | 4.7 | Py88Mo4Cu8 | 51 ± 2 | 0.547 ± 0.010 | 2.100 ± 0.009 | 3.01 ± 0.57 | 13.0 ± 0.3 |
| 8.8 | 7.0 | Py86Mo6Cu8 | 68 ± 2 | 0.407 ± 0.009 | 2.104 ± 0.009 | 1.67 ± 0.40 | 16.5 ± 0.4 |
| 8.8 | 9.4 | Py85Mo8Cu7 | 117 ± 3 | 0.273 ± 0.007 | 2.109 ± 0.009 | 0.97 ± 0.33 | 20.9 ± 0.5 |
| 11.7 | 0 | Py90Cu10 | 29 ± 2 | 0.807 ± 0.015 | 2.098 ± 0.011 | 7.25 ± 0.91 | 8.1 ± 0.2 |

## 3. Results

Mo and Cu doping level dependences of magnetization and exchange stiffness in the co-doped Py system are shown in Fig. 4 (a) and (b). Although both magnetization and exchange stiffness show a decrease with increasing doping levels, the former is very monotonic. The magnetization shows a linear relation with doping in the composition range studied. The Cu (Mo) doping level dependences at different Mo (Cu) doping levels are almost parallel with each other, indicating that the Mo and Cu affect the magnetization independently. The rates of magnetization reduction, per each doping level expressed in at. % with respect to total alloy, are 6.4 and 1.5 % for Mo and Cu, respectively, which are greater than the 1 % reduction expected if the NiFe atoms were simply replaced by non-magnetic atoms. These reduction rates are similar to the results obtained from density functional theory calculations for single-element doped of Mo or Cu, in which Mo addition lowers the Ni and Fe site moments and Mo polarizes antiferromagnetically, while Cu addition lowers the Ni site moment [42]. Since Mo doping and Cu doping can be treated as independent of each other for magnetization, we can determine the magnetization of co-doped Py by summing the reduction in magnetization due to Mo and Cu.

The exchange stiffness of co-doped Py decreases more steeply than the decrease in magnetization, similar to that reported for single-element doping Py systems [22,27]. The reduction is more pronounced with Mo doping compared to Cu doping. The slope of the curves showing the Mo (Cu) dependence of exchange stiffness at each Cu (Mo) doping level varies somewhat depending on the Cu (Mo) doping level. It indicates that the two dopants, Mo and Cu, affect the exchange stiffness in weak correlation with each other, unlike the behavior of magnetization. Here, to clarify the relationship between the magnetization and the exchange stiffness, the exchange stiffness is plotted as a function of magnetization in Fig. 5 for all compositions examined. Surprisingly, the marks representing the $A$-$M_s$ relationship all lie on a single curve expressed in a power, regardless of doping levels of each element. The power exponent is found to be 2.24 ± 0.04. In order to examine the dependence in more detail, power-law fitting is performed for each Cu doping level. The results are shown in the first half of Table 2. The exponents of the power law for each Cu doping level are consistent within the error, indicating that they can still be approximated by a single curve represented by a power. We will discuss this relationship in the next section.

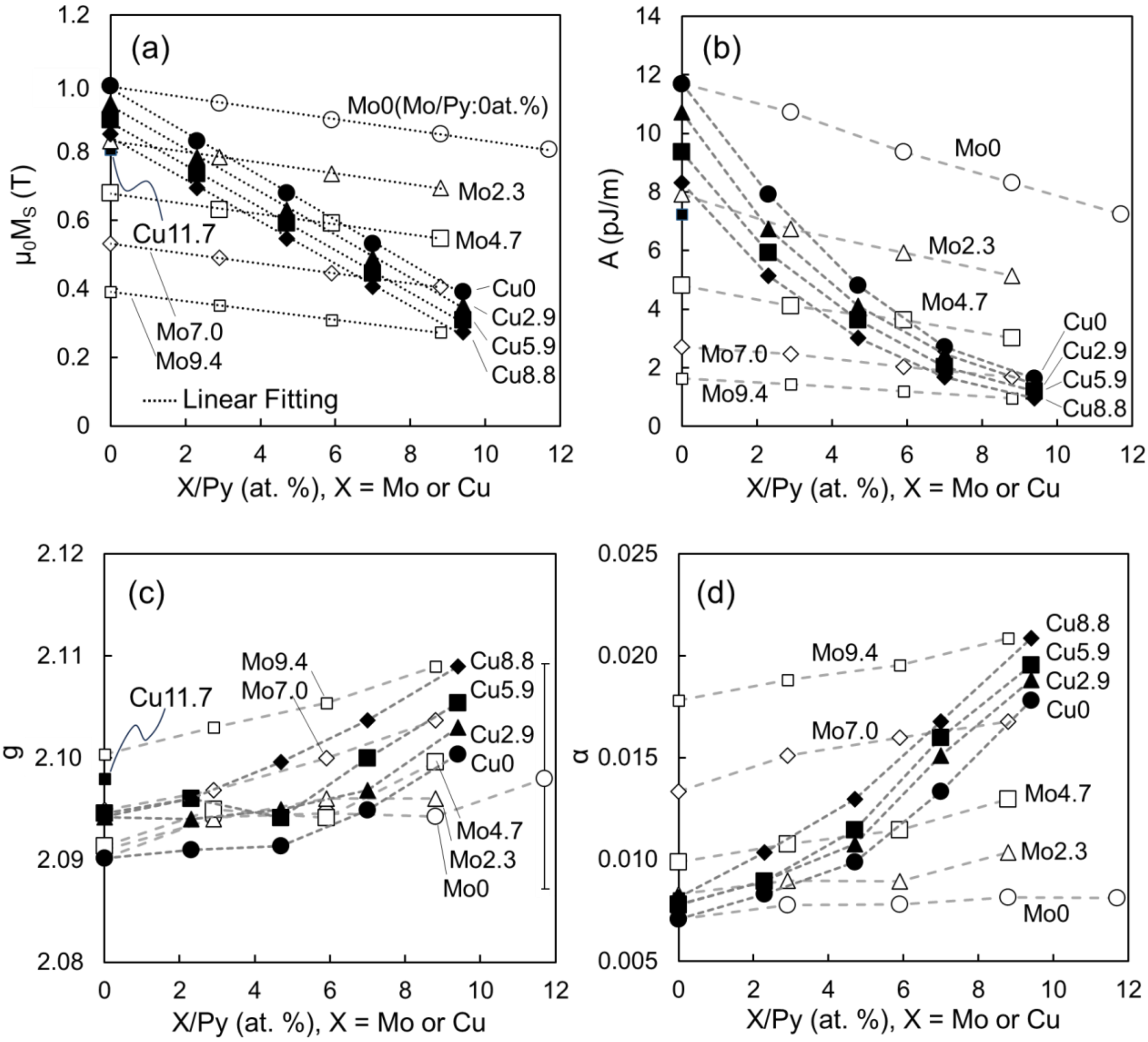

Figure 4. Doping level dependence of magnetization (a), exchange stiffness (b), *g*-factor (c), and damping constant (d) for Py co-doped with Mo and Cu. Black-filled marks indicate the dependence on Mo concentration at constant Cu concentration, with closed circles, closed triangles, closed large squares, closed diamonds, and closed small squares corresponding to Cu doping levels of 0, 2.9, 5.9, 8.8, and 11.7 at. % in Cu/Py, respectively. White-filled marks indicate the dependence on Cu concentration at constant Mo concentration, with open circles, open triangles, open large squares, open diamonds, and open small squares corresponding to the Mo doping levels of 0, 2.3, 4.7, 7.0, and 9.4 at. % in Mo/Py, respectively. The dotted lines in (a) show the results of the linear approximation. Error bars for Figs. 4 (a), (b) and (d) (summarized in Table 1) are omitted. On the other hand, for Fig. 4 (c), where the estimated errors are large, an error bar is marked at a representative point to indicate uncertainty. All plots in Fig. 4 (c) have similar error bars with an average of ±0.010.

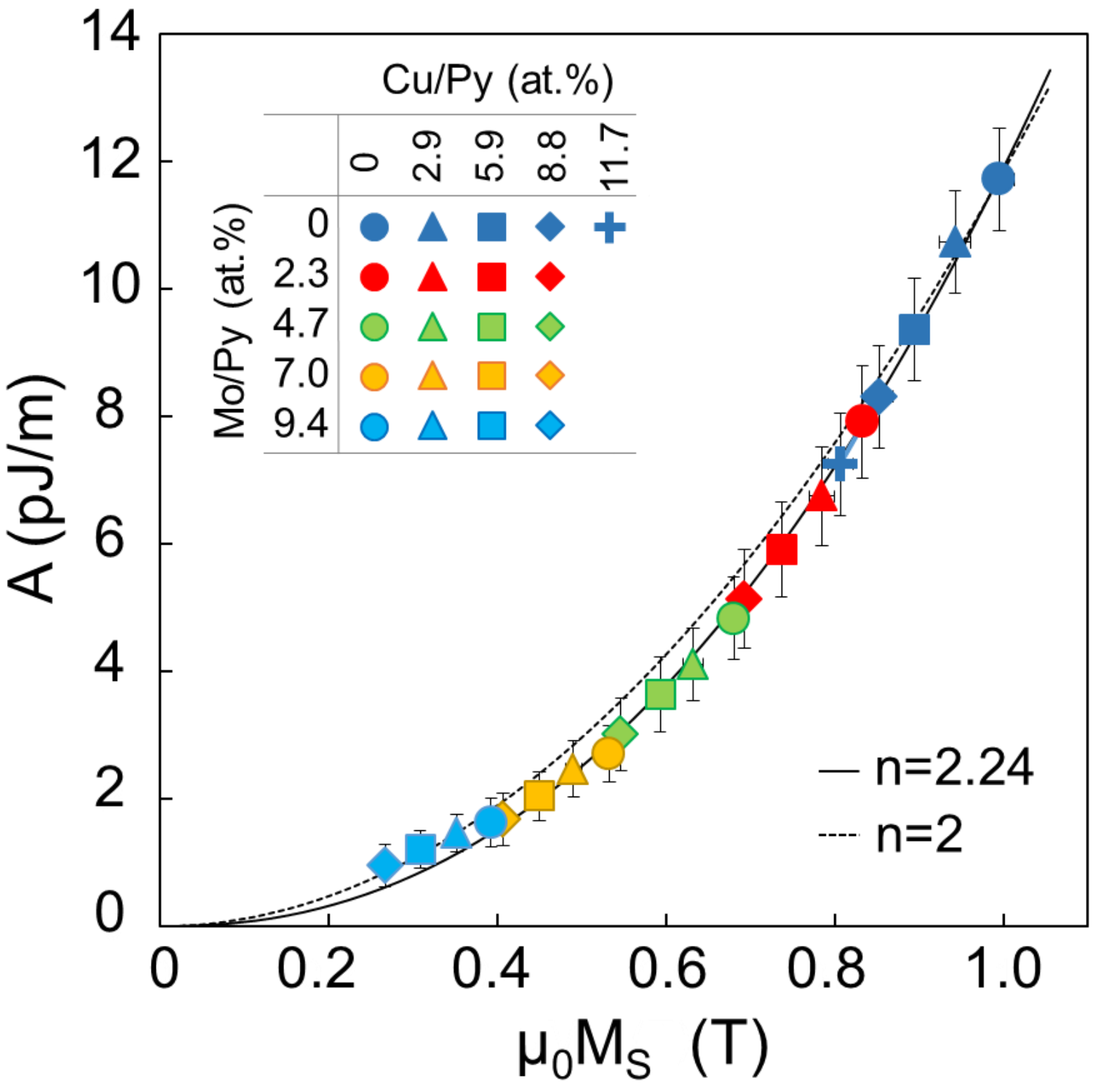

Figure 5. Exchange stiffness as a function of magnetization obtained from Py co-doped with Mo and Cu. Color of the symbol indicates Mo doping level and shape indicates Cu doping level. The solid line is a fit to the power low and the broken line is a fit to square of magnetization.

The dependences of *g*-factor and damping constant $\alpha$ on Mo and Cu doping levels in co-doped systems are shown in Fig. 4 (c) and (d). Note here that the error bar in Fig. 4 (c) is large. Unlike the decreasing trends in the magnetization and the exchange stiffness mentioned above, doping appears to slightly increase the *g*-factor, albeit within the margin of error, and it increases damping, especially for Mo doping. If doping increases the *g*-factor above 2, it implies an increase in the orbital magnetic moment. The increase in orbital moment leads to an increase in the damping through the interaction between spin-orbit interaction (SOI) and spin angular momentum [43, 44]. Thus, one would expect the behavior of *g*-factor to be similar to that of damping. What is characteristic of the damping behavior is that the increase in damping is kept very small for Cu doping compared to Mo doping. It is the electrons near Fermi energy that contribute to the damping [40]. Since the Fermi energy of dopant Cu is occupied by s-electrons which is less susceptible to the SOI, and SOI itself is smaller in lighter Cu than heavier Mo, the damping is considered to be less susceptible to Cu doping.

## 4. Discussion

In the previous section, we have shown from the experimental results that for two-element doped Py, the exchange stiffness can be expressed as a power of the magnetization. Based on the classical Heisenberg model considering the nearest neighbor spins, its Hamiltonian can be written as

$$H = -J \sum_{i,j} \vec{S_i} \cdot \vec{S_j}, \tag{3}$$

with $\vec{S_i}$ being a spin for atom *i* and *J* is the exchange constant between the nearest neighbor spins. Assuming long wave-length deviations and continuum limit, the energy necessary to produce the deviation from the uniformly aligned mode in three-dimensional lattice is derived as follows [2,3,5],

$$\Delta H \propto Ja^2 M_s^2 \left( |\nabla M_x|^2 + \left|\nabla M_y\right|^2 + |\nabla M_z|^2 \right) \tag{4}$$

The prefactor on the right-hand side corresponds to the exchange stiffness, thus $A \propto Ja^2 M_s^2$. If the change in the lattice constant *a* is almost negligible, as in the present case, $A \propto JM_s^2$. Strictly speaking, this model applies to localized electron systems. In itinerant systems such as in this study, the interactions not only with the nearest-neighbor atoms but also with the atoms further away can affect the exchange stiffness [45]. Nevertheless, the trend of the exchange stiffness such as dopant concentration dependence can still be explained by the interactions between the nearest neighbors [33].

For impurity-doped systems, the relation $A \propto Ja^2 M_s^2$ was first suggested by Ref. 22 in order to explain the experimentally obtained steeper decrease in the exchange stiffness than the magnetization with increasing Cu concentration in Py-Cu system. We have investigated the magnetization dependence of the exchange stiffness for some single-element doped Py systems, based on the disclosed values of $A$ and $M_s$ in literatures for doped Py [22,25,33]. The results are summarized in the second half of Table 2. It shows that the $A$ - $M_s$ relationship can be fitted approximately by power and the exponents ranged from 2 to just over 3. The power law therefore seems to be a general trend in Py systems with non-magnetic single-element doping. Interestingly, our results show that the power law holds even for co-doped systems.

Table 2. Exponents derived from fitting the $M_s$ dependence of $A$ to the power law and the coefficients of determination of the fitting.

| Composition | Method | $n$ in $A = c \times M_s^n$ | Coefficient of Determination $R^2$ | Reference |
|---|---|---|---|---|
| (Ni80Fe20)-Mo-Cu | broadband FMR | 2.24 ± 0.04 | 0.997 | This study |
| (Ni80Fe20)-Mo-Cu0% | | 2.28 ± 0.06 | 0.999 | |
| (Ni80Fe20)-Mo-Cu2.9% | | 2.29 ± 0.11 | 0.996 | |
| (Ni80Fe20)-Mo-Cu5.9% | | 2.20 ± 0.10 | 0.997 | |
| (Ni80Fe20)-Mo-Cu8.8% | | 2.19 ± 0.10 | 0.997 | |
| (Ni80Fe20)-Pt | broadband FMR | 2.31 ± 0.42 | 0.881 | 25 |
| (Ni80Fe20)-Au | | 1.94 ± 0.18 | 0.974 | |
| (Ni80Fe20)-Ag | | 2.31 ± 0.16 | 0.988 | |
| (Ni80Fe20)-Cu | broadband FMR | 3.39 ± 0.49 | 0.962 | 22 |
| (Ni81Fe19)-V | *ab initio* calculations | 2.76 ± 0.04 | 1.000 | 33 |
| (Ni81Fe19)-Au | | 2.16 ± 0.02 | 0.999 | |

The exponents of the power law obtained from the doped Py were not 2 but close to 2, predicted by the simple Heisenberg model. Because it is the itinerant electron system, deviations from the prediction will occur and we need to modify the model. In addition, $J$ defined in Equation (3) can also vary depending on the doping elements and their amounts, resulting in the deviation of the exponents. However, since the magnetization dependence of the exchange stiffness of co-doped system exhibits a single curve despite the presence of two doping elements, the change in $J$ may not be so pronounced. The result obtained in this study indicates that magnetization is a dominant factor for exchange stiffness in the doped Py system. We also need to take into considerations the temperature dependence of the magnetization and the exchange stiffness [4,5,46-48]. In Fig. 5, small deviations from the fitting curve are observed at small exchange stiffness and small magnetization region. The decrease in the exchange stiffness may have resulted in a lower Curie temperature, which in turn may have resulted in a smaller magnetization.

This study has dealt with non-magnetic dopants. For magnetic dopants, the power law is not suitable. It has been reported that the doping of Gd, one of the magnetic elements, decreases the exchange stiffness whereas it increases the magnetization [33]. The power law would apply to non-magnetic doping elements and, moreover, to a range of doping levels that do not significantly affect the exchange interaction itself, which is also a useful range in practical permalloys.

## 5. Conclusions

In this study, the effect of Mo and Cu co-doping on the exchange stiffness of Py was investigated using broadband FMR spectroscopy, along with the effect of co-doping on other typical magnetic properties. Mo and Cu co-doping reduces the magnetization of Py linearly with respect to each doping levels and affects magnetization independently of each other, irrespective of co-doping. The rates of magnetization reduction per doping level were 6.4 % and 1.5 % for Mo and Cu, respectively, which are greater than 1% for simply substituting non-magnetic element. For the exchange stiffness, the two dopants lead to a more rapid decrease in the exchange stiffness with a weak correlation with each other. In contrast to these decreasing trends, the *g*-factor appears to have increased slightly, although within the margin of error, and damping was increased, especially with Mo doping. We found that the magnetization dependence of exchange stiffness for all the co-doped Py samples lie on a single curve presented by $A \propto M_s^n$, with the exponent $n$ of 2.24 ± 0.04. The relationship that the exchange stiffness is proportional to the power of magnetization, is also found in the previously reported non-magnetic single-element doped Py systems. The result obtained in the present study that the exchange stiffness can be expressed by a single power law even in the presence of two doping elements indicates that magnetization is the dominant factor in exchange stiffness. The feature that the exchange stiffness is proportional to the power of magnetization remains phenomenological but will provide guidance for the unknown exchange stiffness of non-magnetic element doped systems, facilitating the prediction of the performance of magnetic devices using those systems.

## Acknowledgment

We thank T. Kondo for useful discussions.